\documentclass[10pt,english]{article}
\usepackage{graphicx}
\usepackage[T1]{fontenc}
\usepackage[latin9]{inputenc}
\usepackage[letterpaper]{geometry}
\usepackage{float}
\usepackage{amsbsy}
\usepackage{setspace}
\usepackage{amssymb}
\usepackage{esint}
\makeatletter

\floatstyle{ruled}
\newfloat{algorithm}{tbp}{loa}
\floatname{algorithm}{Algorithm}

\usepackage{babel}

\begin{document}

\title{Monte-Carlo Simulation of a Multi-Dimensional Switch-Like Model of Stem Cell Differentiation}

\author{M. Andrecut}

\maketitle
{

\centering ISIS, University of Calgary, Alberta, T2N 1N4, Canada

}

\section{Introduction}

The process controlling the diferentiation of stem, or progenitor,
cells into one specific functional direction is called lineage specification.
An important characteristic of this process is the multi-lineage priming,
which requires the simultaneous expression of lineage-specific genes.
Prior to commitment to a certain lineage, it has been observed that
these genes exhibit intermediate values of their expression levels.
Multi-lineage differentiation has been reported for various progenitor
cells \cite{hu1997,akashi2003,kim2005,miyamoto2002,swiers2006,loose2006,patient2007,graf2002}, and it has been
explained through the bifurcation of a metastable state \cite{roeder2006,huang2007,chickarmane2009}. 
During the differentiation process the dynamics of the core regulatory network follows a bifurcation,
where the metastable state, corresponding to the progenitor cell,
is destabilized and the system is forced to choose between the possible
developmental alternatives. While this approach gives a reasonable
interpretation of the cell fate decision process, it fails to explain
the multi-lineage priming characteristic. Here, we describe a new
multi-dimensional switch-like model that captures both the process
of cell fate decision and the phenomenon of multi-lineage priming.
We show that in the symmetrical interaction case, the system exhibits
a new type of degenerate bifurcation, characterized by a critical
hyperplane, containing an infinite number of critical steady states.
This critical hyperplane may be interpreted as the support for the
multi-lineage priming states of the progenitor. Also, the cell fate
decision (the multi-stability and switching behavior) can be explained
by a symmetry breaking in the parameter space of this critical hyperplane.
These analytical results are confirmed by Monte-Carlo simulations
of the corresponding chemical master equations.

\section{Stem Cell Differentiation}

\begin{figure}
\centering
\includegraphics[width=7cm]{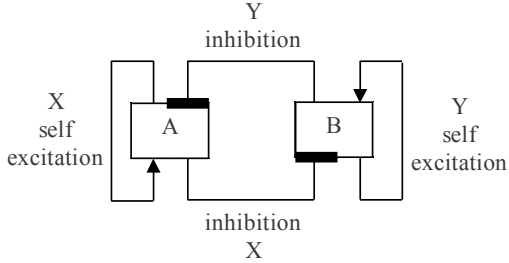}
\caption{The architecture of the self-excitation and inhibition mechanisms in the binary cell fate decision circuit.}
\label{fig1}
\end{figure}

The processes describing the interactions in systems like transcriptional
regulatory networks are extremely complex. Genes can be turned on
or off by the binding of proteins to regulatory sites on the genome
\cite{ozbundak2002, ptashne2002}. The proteins are known as transcription
factors, while the DNA-binding sites are known as promoters. Transcription
factors can regulate the production of other transcription factors,
or they can regulate their own production. The transcription process
can be described by a sequence of reactions, in which RNA polymerase
($R$) binds to a gene's promoter leading to the transcription
of a complete messenger RNA molecule. The genetic information transcribed
into messenger RNA molecules is then translated into proteins by ribosomes.
Thus, the general assumption is that the genes can be excited or inhibited
by the products of the other genes in the network, generating complex
behavior like multi-stability and switching between different steady
state attractors. Based on these general assumptions, it has been
shown that a simple gene regulatory circuit (Fig. \ref{fig1}) in which two
transcription factors, $X$ and $Y$, inhibit each other, and in the
same time activate themselves, can be used as a model of binary cell
fate decision in multipotent stem or progenitor cells \cite{roeder2006,huang2007,chickarmane2009}. 
This circuit can generate multistability
and explains the symmetric precursor state, in which both factors
are present in the cell at equal (low) amounts. This circuit typically
produces three stable \textit{attractor} states that correspond to
observable cell states. The state 1, with the expression pattern $X\gg Y$,
and the state 2, with the opposite pattern $Y\gg X$ represent the
cell fates, while the state 3, with a balanced expression $X\simeq Y$
represents the undecided multipotent state. This simple model provides
a conceptual framework for understanding cell fate decisions, and
it will be used as a starting point in the development of our model.

\section{Monte-Carlo Simulation Approach}

The Monte-Carlo simulation approach employed here is based on the
well known Gillespie algorithm \cite{gillespie1977}, which is a variety
of a dynamic Monte Carlo method. The traditional continuous and deterministic
description of biochemical rate equations, modeled as a set of coupled
ordinary differential equations, relies on bulk reactions that require
the interactions of millions of molecules. In contrast, the Gillespie
stochastic algorithm simulates every reaction explicitly, and calculates
the time evolution of the system by determining the probabilities
of each discrete chemical reaction and the resulting changes in the
number of each molecular species presented in the system. This algorithm
has rigorous theoretical foundations, and gives the exact solution
for a system of elementary chemical reactions in the approximation
of a well-mixed environment. When simulated, a Gillespie realization
represents a random walk that exactly represents the distribution
of the chemical master equation. The algorithm is computationally
expensive and several modifications have been proposed to speed up
computation, including the next reaction method, tau-leaping, as well
as hybrid techniques where abundant reactants are modeled with deterministic
behavior \cite{gibson2000,rathinam2003,slepoy2008}. These adapted
techniques provide a compromise between computational speed and the
exactitude of the theory behind the algorithm as it connects to the
chemical master equation. Here we use the standard stochastic simulation
algorithm, known as the Gillespie's direct method. The rigorous derivation
of the algorithm has been given elsewhere and it has been shown to
remain "exact" for arbitrary low number of molecules \cite{gillespie1977}. 
 
Consider a system composed of $N$ chemical species $X_{\nu}$ $(\nu=1,...,N)$,
interacting through $M$ reactions $R_{\mu}$ $(\mu=1,...,M)$ in
the cell volume $V$. Every chemical reaction $R_{\mu}$ is characterized by its
stochastic rate constant $k_{\mu}$, which depends on the physical
properties of the molecules taking part in the reaction. The product
$k_{\mu}dt$ is the probability that one elementary reaction $R_{\mu}$
happens in the next infinitesimal time interval $dt$. The main steps
of the Gillespie algorithm consist of: 
\begin{description}
\item [{(a)}] calculating the waiting time $\tau$ for the next reaction
to occur;
\item [{(b)}] determining which reaction $\mu$ in the system actually
will occur. 
\end{description}
These quantities are computed by generating two random numbers according
to the following probability density function:\begin{equation}
P(\tau,\mu)=a_{\mu}\exp(-a_{0}\tau),\end{equation}
where \begin{equation}
a_{\mu}=m_{\mu}k_{\mu},\end{equation}
and \begin{equation}
a_{0}=\sum_{\mu=1}^{M}a_{\mu}.\end{equation}
Here, $m_{\mu}$ is the number of distinct reactant combinations available
for the reaction $R_{\mu}$ at the given state of the system. The
coefficient $a_{\mu}$ is called the propensity of reaction $R_{\mu}$.
Thus, $P(\tau,\mu)$ is the probability that the next reaction will
occur in the infinitesimal time interval $dt$ and that it will be
the $R_{\mu}$ reaction. After determination of $\tau$ and $\mu$,
the numbers of molecules in the system are adjusted according to the
reaction $R_{\mu}$. Also, the time $t$ is advanced to $t+\tau$.
The larger the propensity is, the greater is the chance that a given
reaction will happen in the next step of the simulation. It is worth
noting that there is no constant length for a time-step in the simulation.
The length of each time-step is determined independently in every
iteration, and takes different values depending on the state of the
system. 

The implementation of the Gillespie algorithm is straightforward,
and one can find excellent descriptions of it in the literature \cite{adalsteinsson2004,kierzek2002}. 
Below we give the pseudo-code of the algorithm:

\bigskip

\#Gillespie's direct method

1. Set initial numbers of molecules, set time $t\leftarrow0$;

2. Calculate the propensities, $a_{\mu}$, for all $\mu=1,...,M$;

3. Choose $\mu$ with the probability:\begin{equation}
\Pr(reaction=\mu)=\frac{a_{\mu}}{\sum_{\mu=1}^{M}a_{\mu}};\end{equation}

4. Choose $\tau$ with the probability:\begin{equation}
\Pr(time=\tau)=\left(\sum_{\mu=1}^{M}a_{\mu}\right)\exp\left[-\tau\left(\sum_{\mu=1}^{M}a_{\mu}\right)\right];\end{equation}

5. Change the number of molecules to reflect execution of reaction
$\mu$; 

6. Set $t\leftarrow t+\tau$, and go to step 2.

\section{2-Dimensional Model}

We consider the two gene circuit shown in Figure 1. We will focus
on the elementary processes that must occur, such as the promoter
binding of the transcription factors $X$ and $Y$ to the promoters,
$A$ and $B$, respectively, and the activation and degradation of
transcription factors. Also, we propose a general approach to integrate
the two inputs to each gene, which does not depend on the assumption
of cooperativity or other explicit modeling. In order to provide a
quantitative model of this genetic circuit, we employ a formalism
originally developed for the mean-field description of the stochastic
interactions in transcriptional regulatory networks \cite{andrecut2006,andrecut2008}. 
The promoter binding and unbinding, subsequent self-activation, inhibition,
dissociation and the degradation reactions for $X$, and respectively $Y$, are:
\begin{equation}\label{eq6}
\begin{array}{ccccccccc}
 &  &  & _{k_{AR}}\\
A & + & R & \longrightarrow & A & + & R & + & X
\end{array}
\end{equation}
\begin{equation}
\begin{array}{ccccc}
 &  &  & _{k_{AX}^{+}}\\
A & + & X & \longrightarrow & AX
\end{array}
\end{equation}
\begin{equation}
\begin{array}{ccccc}
 & _{k_{AX}^{-}}\\
AX & \longrightarrow & A & + & X
\end{array}
\end{equation}
\begin{equation}
\begin{array}{ccccccccc}
 &  &  & _{k_{X}^{+}}\\
AX & + & R & \longrightarrow & AX & + & R & + & X
\end{array}
\end{equation}
\begin{equation}
\begin{array}{ccccc}
 &  &  & _{k_{AY}^{+}}\\
A & + & Y & \longrightarrow & AY
\end{array}
\end{equation}
\begin{equation}
\begin{array}{ccccc}
 & _{k_{AY}^{-}}\\
AY & \longrightarrow & A & + & Y
\end{array}
\end{equation}
\begin{equation}
\begin{array}{ccc}
 & _{k_{X}^{-}}\\
X & \longrightarrow & \textrm{\ensuremath{\emptyset}}
\end{array}
\end{equation}
\begin{equation}\label{eq13}
\begin{array}{ccccccccc}
 &  &  & _{k_{BR}}\\
B & + & R & \longrightarrow & B & + & R & + & Y
\end{array}
\end{equation}
\begin{equation}
\begin{array}{ccccc}
 &  &  & _{k_{BY}^{+}}\\
B & + & Y & \longrightarrow & BY
\end{array}
\end{equation}
\begin{equation}
\begin{array}{ccccc}
 & _{k_{BY}^{-}}\\
BY & \longrightarrow & B & + & Y
\end{array}
\end{equation}
\begin{equation}
\begin{array}{ccccccccc}
 &  &  & _{k_{Y}^{+}}\\
BY & + & R & \longrightarrow & BY & + & R & + & Y
\end{array}
\end{equation}
\begin{equation}
\begin{array}{ccccc}
 &  &  & _{k_{BX}^{+}}\\
B & + & X & \longrightarrow & BX
\end{array}
\end{equation}
\begin{equation}
\begin{array}{ccccc}
 & _{k_{BX}^{-}}\\
BX & \longrightarrow & B & + & X
\end{array}
\end{equation}
\begin{equation}
\begin{array}{ccc}
 & _{k_{Y}^{-}}\\
Y & \longrightarrow & \textrm{\ensuremath{\emptyset}}
\end{array}
\end{equation}
Here, $k_{AX}^{+}$, $k_{AX}^{-}$, $k_{BY}^{+}$, $k_{BY}^{-}$,
describe the binding and release rates between the transcription factor
and the promoter element, $k_{AY}^{+}$, $k_{AY}^{-}$, $k_{BX}^{+}$,
$k_{BX}^{-}$ correspond to the cross inhibition rates, while $k_{X}^{+}$,
$k_{X}^{-}$, $k_{Y}^{+}$, $k_{Y}^{-}$ reflect the activation and
the degradation rates of the transcription factors. We assume that
the role of the first reaction for each transcription factor, Equation \ref{eq6}
and respectively Equation \ref{eq13}, is just to provide a small "basic level
of expression" (with the rates $k_{AR}$, and respectively $k_{BR}$),
in order to avoid their complete extinction. It's effect is equivalent
to a positive noise term $\eta_{X,Y}$ in the differential equation
describing the dynamics of the transcription factor. Therefore, in
the following analysis we will neglect the contribution of this reaction,
since it doesn't really have an influence on the "logic functionality"
of the circuit. 

The dynamical behavior (rate of change of active levels of the proteins)
of the isolated transcription factors is therefore described by the
stochastic differential equations:\begin{equation}
\frac{d}{dt}[X]=k_{X}^{+}[AX]-k_{X}^{-}[X]+\eta_{X},\end{equation}
\begin{equation}
\frac{d}{dt}[Y]=k_{Y}^{+}[BY]-k_{Y}^{-}[Y]+\eta_{Y},\end{equation}
where $[.]$ denotes concentration. Assuming that the reversible binding-unbinding
processes are in equilibrium, we have:\begin{equation}
k_{AX}[A][X]=[AX],\end{equation}
\begin{equation}
k_{BY}[B][Y]=[BY],\end{equation}
\begin{equation}
k_{AY}[A][Y]=[AY],\end{equation}
\begin{equation}
k_{BX}[B][X]=[BX],\end{equation}
 where $k_{AX}=k_{AX}^{-}/k_{AX}^{+}$, $k_{BY}=k_{BY}^{-}/k_{BY}^{+}$,
$k_{AY}=k_{AY}^{-}/k_{AY}^{+}$, $k_{BX}=k_{BX}^{-}/k_{BX}^{+}$.
Also, since the promoters can be in three different states we have:\begin{equation}
[AX]+[AY]+[A]=[A_{0}],\end{equation}
\begin{equation}
[BY]+[BX]+[B]=[B_{0}],\end{equation}
where $[A_{0}]$ and $[B_{0}]$ are the total concentrations of the
two promoters. From the above equations, and neglecting the noise
terms, we obtain the following system of deterministic differential
equations:\begin{equation}
\frac{d}{dt}x=\alpha x\left(\frac{a_{3}}{a_{1}x+a_{2}y+1}-1\right),\end{equation}
\begin{equation}
\frac{d}{dt}y=\beta y\left(\frac{b_{3}}{b_{1}x+b_{2}y+1}-1\right),\end{equation}
 where we assumed that: $x=[X]$, $y=[Y]$, $\alpha=k_{X}^{-}$, $\beta=k_{Y}^{-}$,
$a_{1}=k_{AX}$, $b_{1}=k_{BX}$, $a_{2}=k_{AY}$, $b_{2}=k_{BY}$,
$a_{3}=[A_{0}]k_{AX}k_{X}^{+}/k_{X}^{-}$, $b_{3}=[B_{0}]k_{BY}k_{Y}^{+}/k_{Y}^{-}$.

We are interested in the symmetrical case, where $\alpha=\beta$,
$a_{1}=b_{2}$, $a_{2}=b_{1}$, $a_{3}=b_{3}$, such that the system
becomes:\begin{equation}
\frac{d}{dt}x=\alpha x\left(\frac{a_{3}}{a_{1}x+a_{2}y+1}-1\right),\end{equation}
\begin{equation}
\frac{d}{dt}y=\alpha y\left(\frac{a_{3}}{a_{2}x+a_{1}y+1}-1\right).\end{equation}
The steady states of the above differential system of equations are
given by the solutions of the non-linear system: \begin{equation}
\frac{d}{dt}x=0\Leftrightarrow F(x,y,\alpha,\{a\})=\alpha x\left(\frac{a_{3}}{a_{1}x+a_{2}y+1}-1\right)=0,\end{equation}
\begin{equation}
\frac{d}{dt}y=0\Leftrightarrow G(x,y,\alpha,\{a\})=\alpha y\left(\frac{a_{3}}{a_{2}x+a_{1}y+1}-1\right)=0.\end{equation}
In this case, one can easily verify that the system has four steady
states:
\begin{equation}
(x_{0},y_{0})=(0,0),
\end{equation}
\begin{equation}
(x_{1},y_{1})=\left(\frac{a_{3}-1}{a_{1}},0\right),
\end{equation}
\begin{equation}
(x_{2},y_{2})=\left(0,\frac{a_{3}-1}{a_{1}}\right),
\end{equation}

\begin{equation}
(x_{3},y_{3})=\frac{1}{a_{1}+a_{2}}\left(a_{3}-1,a_{3}-1\right),\end{equation}
 corresponding to the extinction, exclusive and coexistence equilibria.
These fixed points are positively defined if $a_{3}>1$.

In order to evaluate the local stability we calculate the eigenvalues,
$\lambda$ and $\mu$, of the Jacobian matrix at these steady states:\begin{equation}
J(x,y,\alpha,\{a\})=\left[\begin{array}{cc}
\frac{\partial F}{\partial x} & \frac{\partial F}{\partial y}\\
\frac{\partial G}{\partial x} & \frac{\partial G}{\partial y}\end{array}\right]=\alpha\left[\begin{array}{cc}
\frac{a_{3}(a_{2}y+1)}{(a_{1}x+a_{2}y+1)^{2}}-1 & -\frac{a_{3}a_{2}x}{(a_{1}x+a_{2}y+1)^{2}}\\
-\frac{a_{3}a_{2}y}{(a_{2}x+a_{1}y+1)^{2}} & \frac{a_{3}(a_{2}x+1)}{(a_{2}x+a_{1}y+1)^{2}}-1\end{array}\right].\end{equation}
The eigenvalues of the Jacobian for the extinction state $(x_{0},y_{0})$
are:\begin{equation}
\lambda=\alpha(a_{3}-1)>0,\end{equation}
\begin{equation}
\mu=\alpha(a_{3}-1)>0.\end{equation}
 Thus, this steady state is always unstable, since $a_{3}>1$. The
eigenvalues for the exclusive steady states, $(x_{1},y_{1})$ and
$(x_{2},y_{2})$, are:\begin{equation}
\lambda=-\alpha\frac{a_{3}-1}{a_{3}},\end{equation}
\begin{equation}
\mu=-\alpha\frac{(a_{3}-1)(a_{2}-a_{1})}{a_{1}+a_{2}(a_{3}-1)},\end{equation}
and respectively:\begin{equation}
\lambda=-\alpha\frac{(a_{3}-1)(a_{2}-a_{1})}{a_{1}+a_{2}(a_{3}-1)},\end{equation}
\begin{equation}
\mu=-\alpha\frac{a_{3}-1}{a_{3}}.\end{equation}
Therefore, the exclusive equilibria are stable if $a_{2}>a_{1}$,
and unstable if $a_{2}<a_{1}$. In contrast, the eigenvalues for the
coexistence equilibrium $(x_{3},y_{3})$ are:\begin{equation}
\lambda=-\alpha\frac{(a_{3}-1)(a_{1}+a_{2})}{a_{3}(a_{1}+a_{2})},\end{equation}
\begin{equation}
\mu=-\alpha\frac{(a_{3}-1)(a_{1}-a_{2})}{a_{3}(a_{1}+a_{2})}.\end{equation}
Since $\lambda<0$, this steady state is stable if $\mu<0$, and it
looses stability if $\mu>0$. One can easily see that the stability
condition, $\mu<0$, is equivalent to $a_{2}<a_{1}$. Thus, a change
in the ratio $\rho=a_{1}/a_{2}$, triggers a bifurcation from one
stable steady state $(x_{3},y_{3})$, when $\rho<1$, to two stable
steady states $(x_{1},y_{1})$ and $(x_{2},y_{2})$, when $\rho>1$. 

\begin{figure}
\centering
\includegraphics[width=7.5cm]{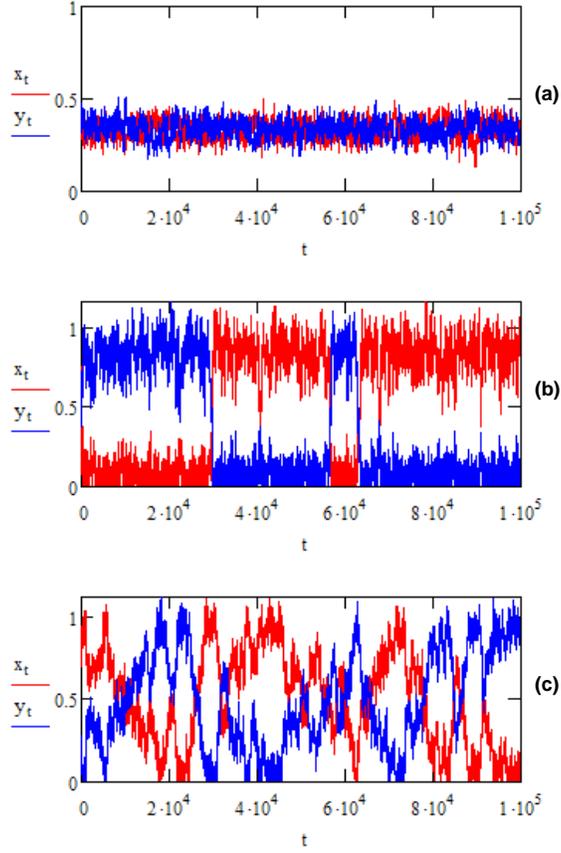}
\caption{Monte-Carlo simulation of the 2-dimensional circuit: (a) $\rho<1$,
one attractor ($k_{AX}^{-}=k_{BY}^{-}=0.5$ and $k_{AY}^{-}=k_{BX}^{-}=1$);
(b) $\rho>1$, two attractors ($k_{AX}^{-}=k_{BY}^{-}=1$ and $k_{AY}^{-}=k_{BX}^{-}=0.5$);
(c) $\rho=1$, the critical case of the degenerate bifurcation ($k_{AX}^{-}=k_{BY}^{-}=k_{AY}^{-}=k_{BX}^{-}=1$).}
\label{fig2}
\end{figure}

\begin{figure}
\centering
\includegraphics[width=7.5cm]{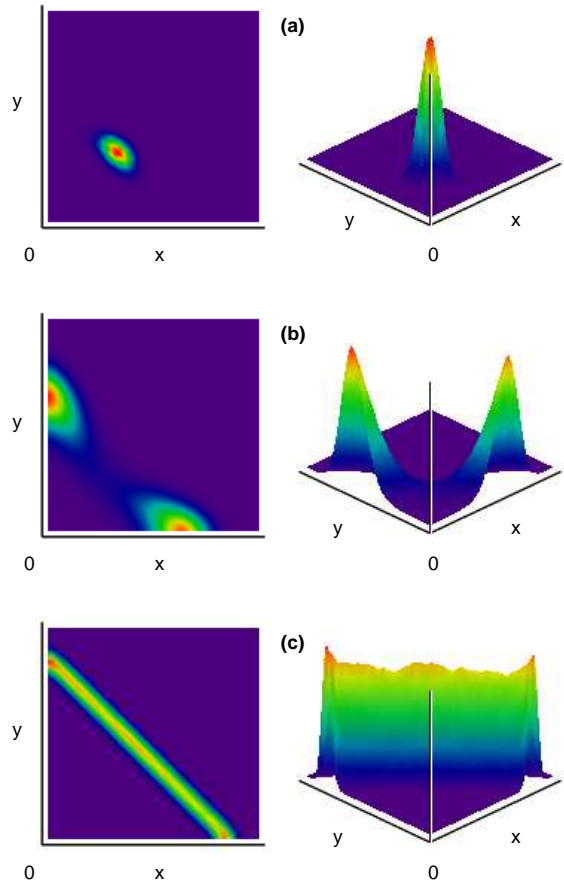}
\caption{The probability density distribution, $P(x,y)$, of the stochastic
trajectory of the 2-dimensional circuit: (a) $\rho<1$, one attractor;
(b) $\rho>1$, two attractors; (c) $\rho=1$, the critical case of
the degenerate bifurcation.}
\label{fig3}
\end{figure}

In Fig. \ref{fig2} and Fig. \ref{fig3} we give the results of the Monte-Carlo simulations.
The initial concentrations and the main reaction constants are set
as: $R=100$, $A_{0}=B_{0}=1$, $X_{0}=Y_{0}=0$, $k_{X}^{-}=k_{Y}^{-}=0.01$,
$k_{A}=k_{B}=0.01$, $k_{X}^{+}=k_{Y}^{+}=0.01$, $k_{AX}^{+}=k_{BY}^{+}=1$,
$k_{AY}^{+}=k_{BX}^{+}=1$. Fig. \ref{fig2} gives the trajectories $x(t)=X(t)/R$
and $y(t)=Y(t)/R$ for $\rho<1$, when there is one noisy attractor,
corresponding to the coexistence equilibrium, $(x_{3},y_{3})$ (Fig. \ref{fig2}(a)), 
and for $\rho>1$, when there are two noisy attractors, corresponding
to the exclusive equilibria, $(x_{1},y_{1})$ and $(x_{2},y_{2})$
(Fig. \ref{fig2}(b)). Also, in Fig. \ref{fig3} we have represented graphically the
probability density distribution, $P(x,y)$, of the transcription
factors (obtained by averaging over $M=10^{4}$ trajectories with
$T=10^{7}$ reactions events). One can see that for $\rho<1$, the
system has only one noisy attractor, corresponding to the stable fixed
point $(x_{3},y_{3})$ (Fig. \ref{fig3}(a), $a_{1}=1$, $a_{2}=2$), while
for $\rho>1$, the system exhibits two noisy attractors corresponding
to the stable fixed points $(x_{1},y_{1})$, and respectively $(x_{2},y_{2})$
(Fig. \ref{fig3}(b), $a_{1}=2$, $a_{2}=1$). We should note that the absolute
values of the rate constants do not play a critical role in the simulation,
as long as their ratios satisfy the bifurcation constraints. 

An important case of the above analysis corresponds to the critical
bifurcation parameter $\rho=1$. In this case the system has the form:\begin{equation}
\frac{d}{dt}x=\alpha x\Phi(x,y,\{a\}),\end{equation}
\begin{equation}
\frac{d}{dt}y=\alpha y\Phi(x,y,\{a\}),\end{equation}
 where\begin{equation}
\Phi(x,y,\{a\})=\frac{a_{3}}{a_{1}(x+y)+1}-1.\end{equation}
One can easily verify that in this case, the exclusive and coexistence
equilibria disappear, and the system has an infinite number of stedy
states $\Omega=\{(x,y)\in\mathbb{R}^{2}|\Phi(x,y,\{a\})=0\}$, which
are practically equivalent to the positive segment of the linear equation:
$x+y=(a_{3}-1)/a_{1}$. These steady states have the following eigenvalues:\begin{equation}
\lambda=0,\end{equation}
\begin{equation}
\mu=-(a_{3}-1)/a_{3}<0.\end{equation}
Therefore, the steady states $\Omega$ are stable, and the system
undergoes a degenerate bifurcation (see Appendix). This situation
is presented in Fig. \ref{fig2}(c) and Fig. \ref{fig3}(c), for $a_{1}=a_{2}=1$. 
One can see that the stochastic system is "undecided", exploring every
point of the critical line with non-zero probability. The line is
attracting, except along itself, that is, there is no "longitudinal" 
force on this line. Therefore every state on it is indifferently stable.
Thus, the critical line becomes an ergodic attractor.

\section{\textit{N}-Dimensional Model}

We consider a $N-$gene circuit, where we denote by $X_{n}$ the transcription
factors and by $A_{n}$ the promoters, $n=1,...,N$. For each gene
we assume the following set of equations:\begin{equation}
\begin{array}{cccccccc}
 &  &  & _{k_{n}}\\
A_{n} & + & R & \longrightarrow & A_{n} & + & R+ & X_{n}\end{array}\end{equation}

\begin{equation}
\begin{array}{ccccc}
 &  &  & _{k_{nm}^{+}}\\
A_{n} & + & X_{m} & \longrightarrow & A_{n}X_{m}\end{array}\end{equation}
\begin{equation}
\begin{array}{ccccc}
 & _{k_{nm}^{-}}\\
A_{n}X_{m} & \longrightarrow & A_{n} & + & X_{m}\end{array}\end{equation}
\begin{equation}
\begin{array}{ccccccccc}
 &  &  & _{k_{n}^{+}}\\
A_{n}X_{n} & + & R & \longrightarrow & A_{n}X_{n} & + & R & + & X_{n}\end{array}\end{equation}
\begin{equation}
\begin{array}{ccc}
 & _{k_{n}^{-}}\\
X_{n} & \longrightarrow & \textrm{\ensuremath{\emptyset}}\end{array}\end{equation}
Thus, the promoter $A_{n}$ can bind to any of the $N$ transcription
factors. Thus, the dynamical behavior of the transcription factors
is described by the following system of stochastic differential equations:\begin{equation}
\frac{d}{dt}[X_{n}]=k_{n}^{+}[A_{n}X_{n}]-k_{n}^{-}[X_{n}]+\eta_{X_{n}},\quad n=1,...,N,\end{equation}
where $\eta_{X_{n}}$ is the noise term corresponding to the first
reaction (51). Assuming that the reversible binding-unbinding processes
are in equilibrium, we have:\begin{equation}
k_{nm}[A_{n}][X_{m}]=[A_{n}X_{m}],\quad n,m=1,...,N,\end{equation}
 where $k_{nm}=k_{nm}^{-}/k_{nm}^{+}$. Also, since the promoters
can be in $N+1$ different states we have:

\begin{equation}
\sum_{m=1}^{N}[A_{n}X_{m}]+[A_{n}]=[A_{n}^{0}],\quad n=1,...,M.\end{equation}
where $[A_{n}^{0}]$ are the total concentrations of the promoters.
From the above equations, and neglecting the noise terms, we obtain
the following system of deterministic differential equations:\begin{equation}
\frac{d}{dt}x_{n}=\alpha_{n}x_{n}\left(\frac{\beta_{n}k_{nn}}{\sum_{m=1}^{N}k_{nm}x_{m}+1}-1\right),\quad n=1,...,N.\end{equation}
 where we assumed that: $x=[X_{n}]$, $\alpha_{n}=k_{n}^{-}$, $\beta_{n}=[A_{n}^{0}]k_{n}^{+}/k_{n}^{-}$.

Let us consider now the symmetric case:\begin{equation}
\frac{d}{dt}x_{n}=F_{n}(x_{1},...,x_{N},\alpha,\beta,\kappa,\gamma)=\alpha x_{n}\left(\frac{\beta\kappa}{\kappa x_{n}+\gamma\sum_{m\neq n}x_{m}+1}-1\right),\quad n=1,...,N,\end{equation}
where $\kappa=k_{nn}$ and $\gamma=k_{nm}$ for $m\neq n=1,...,N$.
The steady states corresponds to the solutions of the nonlinear system:\begin{equation}
\frac{d}{dt}x_{n}=0\Leftrightarrow F_{n}(x_{1},...,x_{N},\alpha,\beta,\kappa,\gamma)=0,\quad n=1,...,N.\end{equation}
There are $N+2$ steady states:\begin{equation}
\left(x_{1}^{(0)},...,x_{N}^{(0)}\right)=\left(0,...,0\right),\end{equation}
\begin{equation}
\left(x_{1}^{(i)},...,x_{n}^{(i)},...,x_{N}^{(i)}\right)=\left(0,...,\frac{\beta\kappa-1}{\kappa},...0\right),\quad i=1,...,N,\end{equation}
\begin{equation}
\left(x_{1}^{(N+1)},...,x_{N}^{(N+1)}\right)=\left(\frac{\beta\kappa-1}{\kappa+(N-1)\gamma},...,\frac{\beta\kappa-1}{\kappa+(N-1)\gamma}\right).\end{equation}
Again, these states correspond to extinction, $N-$exclusive and coexisting
equilibria, and they are positively defined if $\beta\kappa>1$. The
stability of these states can be analyzed using the eigenvalues of
the Jacobian matrix:\begin{equation}
J(x_{1},...,x_{N},\alpha,\beta,\kappa,\gamma)=\left[\frac{\partial}{\partial x_{i}}F_{n}(x_{1},...,x_{N},\alpha,\beta,\kappa,\gamma)\right]_{n,i=1,...,N},\end{equation}
where\begin{equation}
\frac{\partial}{\partial x_{i}}F_{n}=\left\{ \begin{array}{ccc}
\frac{\alpha\beta\kappa\left(\gamma\sum_{m\neq n}x_{m}+1\right)}{\left(\kappa x_{n}+\gamma\sum_{m\neq n}x_{m}+1\right)^{2}}-\alpha & if & n=i\\
-\frac{\alpha\beta\kappa\gamma x_{n}}{\left(\kappa x_{n}+\gamma\sum_{m\neq n}x_{m}+1\right)^{2}} & if & n\neq i\end{array}\right..\end{equation}
The eigenvalues of the extinction state are:\begin{equation}
\lambda_{n}=\alpha\left(\beta\kappa-1\right),\quad n=1,...,N,\end{equation}
which means that this steady state is always unstable, since $\beta\kappa>1$.
For the exclusive equilibria the eigenvalues are:\begin{equation}
\lambda_{n}=\left\{ \begin{array}{ccc}
-\frac{\alpha(\beta\kappa-1)}{\beta\kappa} & if & n=i\\
-\frac{\alpha(\beta\kappa-1)(\gamma-\kappa)}{\gamma(\beta\kappa-1)+\kappa} & if & n\neq i\end{array}\right.,\quad n=1,...,N.\end{equation}
Therefore, these states become stable if $\gamma>\kappa$, and unstable
if $\gamma<\kappa$. In the case of coexisting equilibrium the Jacobian
is given by:\begin{equation}
\frac{\partial}{\partial x_{i}}F_{n}=\left\{ \begin{array}{ccc}
\frac{\alpha(1-\beta\kappa)}{\beta[\kappa+(N-1)\gamma]} & if & n=i\\
-\frac{\alpha\gamma(\beta\kappa-1)}{\beta k[\kappa+(N-1)\gamma]} & if & n\neq i\end{array}\right.,\end{equation}
and it has the following eigenvalues:\begin{equation}
\lambda_{n}=\left\{ \begin{array}{ccc}
-\frac{\alpha(\beta\kappa-1)}{\beta\kappa} & if & n=i\\
-\frac{\alpha(\beta\kappa-1)(\kappa-\gamma)}{\beta k[\kappa+(N-1)\gamma]} & if & n\neq i\end{array}\right.,\quad n=1,...,N.\end{equation}
Thus, this state becomes stable if $\gamma<\kappa$, and unstable
if $\gamma>\kappa$.

In the critical case, $\kappa=\gamma$, the steady state equations
are degenerated and we have again an infinite number of steady states,
all of them satisfying the critical hyperplane equation:\begin{equation}
\sum_{n=1}^{N}x_{n}=\frac{\beta\kappa-1}{\kappa}.\end{equation}
In this critical case the Jacobian takes the simplified form:\begin{equation}
\frac{\partial}{\partial x_{i}}F_{n}=-\frac{\alpha}{\beta}x_{n},\end{equation}
and it has the following eigenvalues:\begin{equation}
\lambda_{n}=\left\{ \begin{array}{ccc}
-\frac{\alpha(\beta\kappa-1)}{\beta\kappa} & if & n=i\\
0 & if & n\neq i\end{array}\right.,\quad n=1,...,N.\end{equation}
Thus, one eigenvalue is always negative, since $\beta\kappa>1$, and
the other $N-1$ eigenvalues are zero. Therefore, the hyperplane containing
the infinite number of steady states is attractive and marginally
stable. 

\begin{figure}
\centering
\includegraphics[width=7.5cm]{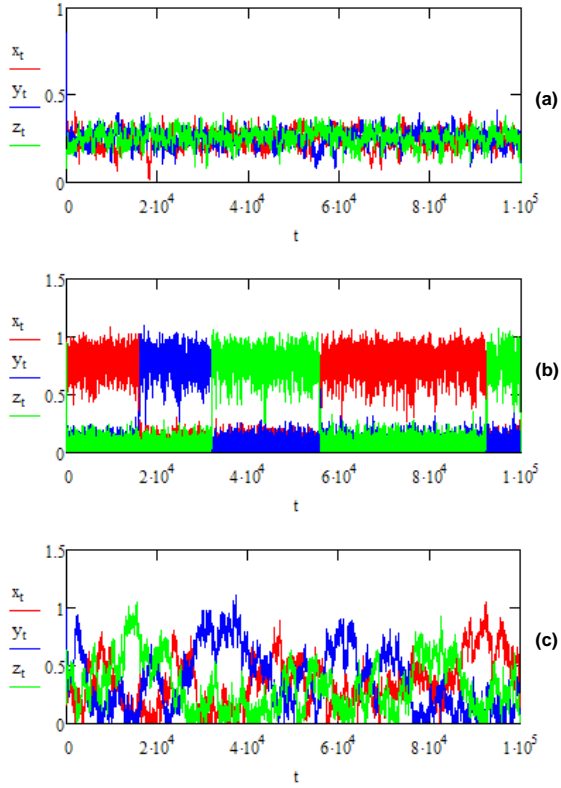}
\caption{Monte-Carlo simulation of the 3-dimensional circuit: (a) $\gamma<\kappa$, one attractor; (b) $\gamma>\kappa$, three attractors; 
(c) $\gamma=\kappa$, the critical case of the degenerate bifurcation.}
\label{fig4}
\end{figure}

In Fig. \ref{fig4} we give the simulation results for a circuit consisting
of three genes, $N=3$. The initial concentrations are set as $R=150$,
$A_{1}=A_{2}=A_{3}=1$, $X_{1}=X_{2}=X_{3}=0$. The rate constants
are the same as for the 2-dimensional circuit. Fig. \ref{fig4} gives the
trajectories $x_{n}(t)$ for $\gamma<\kappa$, when there is one noisy
attractor, corresponding to the coexistence equilibrium, $(x_{1}^{(3)},x_{2}^{(3)},x_{3}^{(3)})$
(Fig. \ref{fig4}(a)), and for $\gamma>\kappa$, when there are three noisy
attractors, corresponding to the exclusive equilibria, $(x_{1}^{(1)},x_{2}^{(1)},x_{3}^{(1)})=\left(\frac{\beta\kappa-1}{\kappa},0,0\right)$,
$(x_{1}^{(2)},x_{2}^{(2)},x_{3}^{(2)})=\left(0,\frac{\beta\kappa-1}{\kappa},0\right)$
and $(x_{1}^{(3)},x_{2}^{(3)},x_{3}^{(3)})=\left(0,0,\frac{\beta\kappa-1}{\kappa}\right)$
(Fig. \ref{fig4}(b)), and in the degenerate case when the plane $x_{1}+x_{2}+x_{3}=\frac{\beta\kappa-1}{\kappa}$
is the ergodic attractor (Fig. \ref{fig4}(c)).

\section{Conclusion}

We have presented a new multi-dimensional switch-like model that captures
both the process of cell fate decision and the phenomenon of multi-lineage
priming. The previous attempts to model the coexistence of three discrete
stable states are based on a complicated molecular interaction mechanisms,
requiring cooperativity or additional transcription factors \cite{roeder2006,huang2007,chickarmane2009}. 
Here, we have shown that very elementary cross-inhibition between two genes and independent autoactivation
can give rise to multistability without cooperativity. It is important
to note the obvious fact that the real molecular mechanisms that govern
the dynamics of this gene regulatory circuit are by orders of magnitudes
more complex, involving perhaps thousands of steps not accounted for
in the presented model. However, this simplified description is still
able to capture to qualitative cell fate decision behavior, specifically,
the existence of an indeterminate multi-potent progenitor state with
equal levels of transcription factors, and the generation of stable
attractor states with asymmetric expression patterns. Also, we have
shown that in the symmetrical interaction case, the system exhibits
a new type of degenerate bifurcation, characterized by a critical
hyperplane containing an infinite number of critical steady states.
This degeneration of the central attractor state captures the intrinsic
heterogeneity of the undecided multipotent state allowing individual
cells to a range of states, and may be interpreted as the support
for the multi-lineage priming states of the progenitor. Also, the
cell fate decision (the multi-stability and switching behavior) can
be explained by a symmetry breaking in the parameter space of this
critical hyperplane. It is important to note here that the critical
hyperplane is also ergodic. Thus, in the critical regime, any stochastic
trajectory of the system will be attracted to the critical hyperplane.
Also, in this case, the dynamics will become confined to this region,
such that the system will visit all the points of the critical hyperplane
with non-zero probability (the priming phenomenon). However, any perturbation
of this critical hyperplane will force the system to collapse in one
of its non-trivial stable steady states (the cell fate decision process).

\section{Appendix: Degenerate Steady State Bifurcation}

We consider a 2-dimensional system of stochastic differential equations
(SDE), of the following generic form:\begin{equation}
(S)\;\left\{ \begin{array}{c}
\dot{x}=\alpha(\bar{x}-x)\Phi(x,y,\{\gamma\})f(x,y,\{a\})+\eta_{x}\\
\dot{y}=\beta(\bar{y}-y)\Phi(x,y,\{\gamma\})g(x,y,\{b\})+\eta_{y}\end{array}\right.,\end{equation}
where $\Phi,f,g:\mathbb{R}^{2}\rightarrow\mathbb{R}$ , $x,y,\bar{x},\bar{y},\alpha,\beta,\{\gamma\},\{a\},\{b\}\in\mathbb{R}$,
and $\eta_{x}$, $\eta_{y}$ correspond to additive noise terms. Also,
we denote by $\Omega=\{(x,y)\in\mathbb{R}^{2}|\Phi(x,y,\{\gamma\})=0\}$
the set of solutions of the equation $\Phi=0$, and we assume that:
$f(x,y,\{a\})\neq0$ and $g(x,y,\{b\})\neq0$, for any $(x,y)\in\mathbb{R}^{2}$.

\textbf{Theorem 1:} The SDE system $(S)$ exhibits a degenerate
bifurcation, $(\bar{x},\bar{y})\rightarrow\widetilde{\Omega}\subset\Omega$,
from one steady state $(\bar{x},\bar{y})$ to a subset of steady states
$\widetilde{\Omega}\subset\Omega$, if: \begin{equation}\label{eq76}
\max\{-\alpha\Phi(\bar{x},\bar{y},\{\gamma\})f(\bar{x},\bar{y},\{a\}),-\beta\Phi(\bar{x},\bar{y},\{\gamma\})g(\bar{x},\bar{y},\{b\}\}>0,\end{equation}
 and\begin{equation}\label{eq77}
\nabla_{v}\Phi(x,y,\{\gamma\})=\left\langle v,\nabla\Phi(x,y,\{\gamma\})\right\rangle <0,\end{equation}
 for any $(x,y)\in\widetilde{\Omega}$ and \begin{equation}
v=[\alpha(\bar{x}-x)f(x,y,\{a\}),\beta(\bar{y}-y)g(x,y,\{b\})]^{T}.\end{equation}

\textbf{Proof:} A first steady state of the system $(S)$
is $(\bar{x},\bar{y})$. Also the system has an infinite number of
steady states $\Omega$, corresponding to the solutions of the equation
$\Phi(x,y,\{\gamma\})=0$. The stability of the steady states can
be analyzed using the eigenvalues, $\lambda_{0}$ and $\lambda_{1}$,
of the Jacobian matrix $J$, which are given by the solutions of the
equation $|J-\lambda I|=0$, where $I$ is the identity matrix. 
In general, the eigenvalues are complex numbers, and
the distance between the solution of the system and thesteady state
changes at an exponential rate, given by the real part of the eigenvalue.
For simplicity the following discussion is restricted to real eigenvalues,
though steady states with complex eigenvalues have similar properties
based on the value of the real part of the eigenvalue. A negative
eigenvalue implies that the solution approaches the steady state along
the corresponding eigenvector, while a positive eigenvalue implies
that the solution moves away from the steady state along the eigenvector.
In a 2-dimensional system there are three possible cases. A stable
steady state has two negative eigenvalues, and hence attracts all
the solutions in a surrounding region. An unstable steady state has
two positive eigenvalues and all the solutions in its neighborhood
move away from it. A saddle point has one negative and one positive
eigenvalue. Now let us analyze the stability of the steady states
of the system. The eigenvalues of the Jacobian for $(\bar{x},\bar{y})$
are:\begin{equation}
\left\{ \begin{array}{c}
\lambda_{0}=-\alpha f(\bar{x},\bar{y},\{a\})\Phi(\bar{x},\bar{y},\{\gamma\})\\
\lambda_{1}=-\beta g(\bar{x},\bar{y},\{b\})\Phi(\bar{x},\bar{y},\{\gamma\})\end{array}\right..\end{equation}
 Thus, this steady state is stable if $\lambda_{0},\lambda_{1}<0$,
and it looses stability if $\lambda_{0}>0$ or $\lambda_{1}>0$, which
is equivalent to the condition imposed by the Equation \ref{eq76}. The other
steady states $(x,y)\in\Omega$, have the following eigenvalues:\begin{equation}
\left\{ \begin{array}{c}
\lambda_{0}=0\\
\lambda_{1}=\alpha(\bar{x}-x)f\frac{\partial\Phi}{\partial x}+\beta(\bar{y}-y)g\frac{\partial\Phi}{\partial y}\end{array}\right.,\end{equation}
 and they are degenerated, since they have at least one zero eigenvalue.
Also, these degenerate steady states become stable for $\lambda_{1}<0$.
Since $\lambda_{1}=\left\langle v,\nabla\Phi\right\rangle =\nabla_{v}\Phi$,
this stability condition is equivalent to the condition imposed by
the Equation \ref{eq77}, which requires that the derivative of $\Phi$ in direction
$v$ must be negative. The directional derivative of a manifold $\Phi$
along a vector $v$ at a given point $(x,y)$, intuitively represents
the instantaneous rate of change of the manifold, moving through $(x,y)$,
in the direction of $v$. One can easily verify that
in this case, $v$ is the eigenvector of the Jacobian corresponding
to the eigenvalue $\lambda_{1}$, that is we have: $Jv=\lambda_{1}v$.
Thus, any change in the parameters $\{\gamma\}$, such that the stable
steady state $(\bar{x},\bar{y})$ becomes unstable, and the steady
states $\widetilde{\Omega}=\{(x,y)\in\Omega|\nabla_{v}\Phi(x,y,\{\gamma\})<0\}$
become stable, results in a degenerate bifurcation of the dynamics
of the stochastic system $(S)$.

A similar property can be formulated for stochastic discrete maps (SDM) 
of the following generic form:\begin{equation}
(M)\;\left\{ \begin{array}{c}
x_{t+1}=(x_{t}-\bar{x})[1-\alpha\Phi(x_{t},y_{t},\{\gamma\})f(x_{t},y_{t},\{a\})]+\bar{x}+\eta_{x}\\
y_{t+1}=(y_{t}-\bar{y})[1-\beta\Phi(x_{t},y_{t},\{\gamma\})g(x_{t},y_{t},\{b\})]+\bar{y}+\eta_{y}\end{array}\right.,\end{equation}
 where $\Phi,f,g:\mathbb{R}^{2}\rightarrow\mathbb{R}$ , $x_{t},y_{t},\bar{x},\bar{y},\alpha,\beta,\{\gamma\},\{a\},\{b\}\in\mathbb{R}$,
and $\eta_{x}$, $\eta_{y}$ correspond to additive noise terms. We
denote by $\Omega=\{(x,y)\in\mathbb{R}^{2}|\Phi(x,y,\{\gamma\})=0\}$
the set of solutions of the equation $\Phi=0$. Also, we assume that:
$f(x,y,\{a\})\neq0$ and $g(x,y,\{b\})\neq0$, for any $(x,y)\in\mathbb{R}^{2}$.

\textbf{Theorem 2:} The SDM system $(M)$ exhibits a degenerate
steady state bifurcation, $(\bar{x},\bar{y})\rightarrow\widetilde{\Omega}\subset\Omega$,
from one steady state $(\bar{x},\bar{y})$ to a subset of steady states
$\widetilde{\Omega}\subset\Omega$, if: \begin{equation}\label{eq82}
\max\left\{ \left|1-\alpha\Phi(\bar{x},\bar{y},\{\gamma\})f(\bar{x},\bar{y},\{a\})\right|,\left|1-\beta\Phi(\bar{x},\bar{y},\{\gamma\})g(\bar{x},\bar{y},\{b\})\right|\right\} >1,\end{equation}
 and \begin{equation}\label{eq83}
\left|1-\nabla_{v}\Phi(x,y,\{\gamma\})\right|<1,\end{equation}
 for any $(x,y)\in\widetilde{\Omega}$ and \begin{equation}
v=[\alpha(x-\bar{x})f(x,y,\{a\}),\beta(y-\bar{y})g(x,y,\{b\})]^{T}.\end{equation}

\textbf{Proof:} The steady states of the map $(P)$ are $(\bar{x},\bar{y})$,
and the set $\Omega$, corresponding to the solutions of the equation
$\Phi(x,y,\{\gamma\})=0$. As before, the eigenvalues, $\lambda_{0}$
and $\lambda_{1}$, are given by the equation $|J-\lambda I|=0$,
where $J$ is the Jacobian of the discrete map $(P)$. In the case
of discrete maps, a stable steady state is characterized by $|\lambda_{0}|<1$
and $|\lambda_{1}|<1$, while an unstable steady state is characterized
by $|\lambda_{0}|>1$ or $|\lambda_{1}|>1$.
The eigenvalues of the Jacobian for $(\bar{x},\bar{y})$ are:\begin{equation}
\left\{ \begin{array}{c}
\lambda_{0}=1-\alpha\Phi(\bar{x},\bar{y},\{\gamma\})f(\bar{x},\bar{y},\{a\})\\
\lambda_{1}=1-\beta\Phi(\bar{x},\bar{y},\{\gamma\})g(\bar{x},\bar{y},\{b\})\end{array}\right..\end{equation}
 Thus, this steady state is stable if $|\lambda_{0}|<1$ and $|\lambda_{1}|<1$,
and it becomes unstable if $|\lambda_{0}|>1$ or $|\lambda_{1}|>1$,
which is equivalent to the condition imposed by the Equation \ref{eq82}. The other
steady states $\Omega$, have the following eigenvalues:\begin{equation}
\left\{ \begin{array}{c}
\lambda_{0}=1\\
\lambda_{1}=1-\alpha(x-\bar{x})f\frac{\partial\Phi}{\partial x}-\beta(y-\bar{y})g\frac{\partial\Phi}{\partial y}\end{array}\right.,\end{equation}
 and they are degenerated, since they have at least one eigenvalue
equal to one. These degenerate steady states become stable for $\left|\lambda_{1}\right|<1$,
which is equivalent to the condition imposed by the Equation \ref{eq83}. Also,
one can verify that $v$ is the eigenvector of the Jacobian, corresponding
to the eigenvalue $\lambda_{1}$. Thus, any change in the parameters
$\{\gamma\}$, such that the stable steady state $(\bar{x},\bar{y})$
becomes unstable, and the steady states $\widetilde{\Omega}=\{(x,y)\in\Omega|\left|1-\nabla_{v}\Phi(x,y,\{\gamma\})\right|<1\}$
become stable, results in a degenerate bifurcation of the dynamics
of the stochastic discrete map $(M)$.


\end{document}